%% file: main.tex
\documentclass[preprint, 12pt]{elsarticle}
\usepackage{amssymb}
\usepackage{amsmath}
\usepackage{float}
\usepackage[flushleft]{threeparttable}
\usepackage{longtable}
\usepackage{rotating}
\usepackage{changepage}
\usepackage{booktabs}
\usepackage{tabularx}
\usepackage{makecell}
\usepackage{microtype}
\usepackage{xurl}

\usepackage{hyperref}
\hypersetup{
    pdftitle={CHIMERA Challenge: Biochemical Recurrence Prediction in Prostate Cancer Patients using multimodal datasets},
    pdfauthor={R.N. Spaans, N. Khalili, C. Chia, T. Wang, G. Litjens},
    pdfkeywords={Prostate Cancer, Artificial Intelligence, Medical Imaging, Deep Learning},
    colorlinks=true, % Optional: makes links colored
    linkcolor=blue,  % Optional
    urlcolor=blue    % Optional
}

\biboptions{sort&compress}
\journal{Medical Image Analysis}

\begin{document}

\begin{frontmatter}

% --- Title, Author, Abstract, etc. ---

\title{CHIMERA Challenge:\\
\large Biochemical Recurrence Prediction in Prostate Cancer Patients using multimodal datasets}

\author[1,2,3]{Robert N. Spaans\corref{mycorrespondingauthor}}
\cortext[mycorrespondingauthor]{Corresponding author}
\ead{robert.spaans@radboudumc.nl}

\author[1,3,4,6]{Catherine Chia}
\author[1,5,6]{Tongjie Wang}
\author[7]{Adam Kowalewski}
\author[8]{Parandzem Khachatryan}
\author[9]{Domingos Oliveira}
\author[1]{Khrystyna Faryna}
\author[2]{Jean-Paul A. van Basten}
\author[1,3]{Geert Litjens}
\author[1]{Nadieh Khalili}
\author[]{on behalf of the CHIMERA Consortium\fnref{consortium}}
\fntext[consortium]{A full list of CHIMERA Consortium members and affiliations is provided at the end of this article.}

\affiliation[1]{organization={Radboud University Medical Center},
            addressline={Department of Pathology, Research Institute for Medical Innovation},
            city={Nijmegen},
            country={The Netherlands}}
\affiliation[2]{organization={Canisius Wilhelmina Hospital},
            addressline={Department of Urology},
            city={Nijmegen},
            country={The Netherlands}}
\affiliation[3]{organization={Oncode Institute},
            city={Utrecht},
            country={The Netherlands}}
\affiliation[4]{organization={Erasmus University Medical Center},
            addressline={Department of Pathology},
            city={Rotterdam},
            country={The Netherlands}}
\affiliation[5]{organization={Erasmus University Medical Center},
            addressline={Department of Urology},
            city={Rotterdam},
            country={The Netherlands}}
\affiliation[6]{organization={Erasmus University Medical Center},
            addressline={Department of Dermatology},
            city={Rotterdam},
            country={The Netherlands}}
\affiliation[7]{organization={Bydgoszcz University of Science and Technology},
            addressline={Faculty of Medicine},
            city={Bydgoszcz},
            country={Poland}}
\affiliation[8]{organization={Yerevan State Medical University},
            city={Yerevan},
            country={Armenia}}
\affiliation[9]{organization={IMP Diagnostics},
            addressline={Research and Development Unit},
            city={Porto},
            country={Portugal}}
            
\begin{abstract}
Biochemical recurrence (BCR), defined as any detectable prostate-specific antigen level after prostatectomy with confirmatory elevation, is widely used as a surrogate endpoint and typically assessed using clinical and pathological variables. Despite advances in multimodal modeling, no standardized benchmark exists for multimodal prognostic modeling in urological cancers, partly because curating heterogeneous multimodal data remains challenging.

We developed the CHIMERA Challenge, a multimodal benchmark integrating preoperative mpMRI, post-prostatectomy histopathology, patient characteristics, and clinician-derived variables from 267 patients across two institutions. The released dataset comprises 801 MRI sequences, 13 clinical variables per case, and 942 WSIs. Training(n=95), validation(n=23), and test(n=149) splits were established and hosted on the Grand Challenge platform. Baseline clinical and pathological characteristics did not differ significantly across splits. Models were evaluated on predicting time to BCR using the C-index. Post-challenge analyses tested how robustly each model type performed when clinician-derived variables were withheld or randomized.

Unimodal clinical models achieved the highest test C-index of 0.7402 but proved highly sensitive to the integrity of these variables, with performance collapsing toward chance ($C$$\approx$0.50) when they were randomized. Multimodal models retained near-baseline performance when these variables were withheld ($\Delta$$C$$\leq$0.04), indicating their ability to recover prognostic signal directly from imaging data.

CHIMERA is the first public, standardized multimodal benchmark for prostate cancer prognosis. Although models using only patient characteristics and clinician-derived variables yielded the highest leaderboard performance, multimodal models combining mpMRI and WSIs demonstrated greater robustness in clinically realistic scenarios where complete expert annotation is not guaranteed.

\end{abstract}


\end{frontmatter}
\newpage
% --- Main text of article ---
\input{sections/introduction.tex} 
\input{sections/method.tex} 
\input{sections/results.tex}
\input{sections/discussion.tex}
\input{sections/conclusion.tex}

% Credits of all authors
\input{sections/credit_section}

% --- Mandatory and Optional Sections Before Bibliography ---
\section*{Declaration of Competing Interest}
G.L. reports grants from the Dutch Cancer Society, the NWO, and the European Union, outside the submitted work. G.L is associate editor for the Medical Image Analysis Journal. All other authors declare no relevant competing interests concerning this paper.

% Acknowledge funding sources and individuals here.
\section*{Acknowledgments}
This work was supported by the Hanarth Fonds and is financed by the sector plan medical sciences of the ministry of the UCW of the Netherlands. The CHIMERA challenge prizes were sponsored by Astellas Pharma, which had no role in the challenge design, data curation, definition of the evaluation metric, assessment of submitted algorithms, or the decision to publish. We thank Jean-Paul A. van Basten for their helpful discussions. The computational resources used by one of the participating teams were enabled in part by support provided by the Digital Research Alliance of Canada (alliancecan.ca).

\input{sections/consortium.tex}
% --- Bibliography ---
\bibliographystyle{elsarticle-num}
\bibliography{references} % Assuming your file is named references.bib

\newpage
\input{sections/supplementary.tex}

\end{document}

%% file: sections/introduction.tex
\section{Introduction}
\label{sec:introduction}

Several benchmark initiatives have demonstrated the value of standardized datasets and fixed evaluation protocols in prognostic modeling in oncology, such as the PANDA challenge for Gleason grading in prostate biopsies \cite{Bulten2022}, the PI-CAI challenge for multiparametric MRI-based (mpMRI) detection of clinically significant prostate cancer \cite{Saha2024}, and the LEOPARD challenge for time-to-recurrence prediction using digitized pathology prostatectomy images \cite{grisi2026leopard, faryna2026leopard}. However, despite recent developments in multimodal model architectures \cite{Marin2025, Lee2023, Schouten2025}, the majority of benchmarks address a task using a single modality (only images or only structured data), leaving open the challenge of prognostic modeling that integrates paired modalities such as raw imaging data and structured data consisting of patient characteristics and/or clinician-derived variables from the same patients.

In prostate cancer specifically, radical prostatectomy (RP) is often advised as curative treatment for patients with intermediate-risk and a life expectancy of $\geq$ 10 years. While the role of prostate-specific antigen (PSA) in initial screening remains debated, it remains an essential biomarker to assess the effectiveness during follow-up care post-RP with concentrations typically falling to undetectable levels (<0.1 ng/mL) within two months after surgery. A subsequent rise in PSA, referred to as biochemical recurrence (BCR), occurs in 27\% to 53\% of patients after undergoing RP. Therefore, risk assessment for BCR is essential for predicting metastasis and mortality risk, enabling specialists to tailor follow-up protocols and salvage treatment decisions to the individual patient \cite{EAUGuidelines2025}.

Established risk models such as CAPRA-S already integrate multimodal perspectives and data to predict BCR after RP such as PSA, Gleason grade, surgical margin status, capsular penetration, seminal vesicle invasion, and lymph node invasion \cite{Cooperberg2011}. Although these variables are represented in tabular form, they are a combination of lab results and clinician-derived information originating from different diagnostic processes.

Building on the foundations established by existing benchmarks, the next step for prognostic AI is to evaluate models in settings that more closely reflect the integrated nature of clinical risk assessment. In prostate cancer, this requires multimodal benchmarks that provide paired raw imaging data, patient characteristics, and clinician-derived variables from the same patients. Such benchmarks enable assessment of whether AI models can learn prognostic information directly from the underlying radiology and pathology data, and whether this information can match, complement, or improve upon structured clinical risk models in a standardized and reproducible setting.

To address this gap, we organized the CHIMERA (Combining HIstology, Medical Imaging (Radiology), and molEcular Data for Medical pRognosis and diAgnosis) challenge. We compiled and released a curated paired multimodal prostatectomy cohort, including radiological and pathological imaging and structured clinical and clinician-derived variables. CHIMERA is a standardized, openly accessible benchmark aimed at prognostic, multimodal AI development in prostate cancer, specifically prediction of time to BCR after radical prostatectomy. Additionally, we performed modality masking experiments to investigate the relative contribution and redundancy of several data sources for single modal models and multimodal fusion models.

%% file: sections/method.tex
\section{Material and methods}
\label{sec:methods}

\subsection{Dataset}
\subsubsection{Inclusion criteria}
The target cohort for the participating algorithms comprises patients undergoing radical prostatectomy for prostate cancer. The challenge cohort consisted of 267 such patients treated at two Dutch centers between 2012 and 2021. Inclusion criteria were prostatectomy between 2012 and 2021, availability of preoperative mpMRI, availability of digitized H\&E-stained prostate WSIs, and no objection to the use of clinical data for medical-scientific research. A full overview of patient demographics, stratified by split, is provided in Table \ref{tab:patient_demographic}.

In addition to the CHIMERA dataset, participants were allowed and encouraged to use publicly available external datasets. Consequently, some participants incorporated the LEOPARD dataset and the Miro-120 dataset \cite{grisi2026leopard,faryna2026leopard,Fenner2025}. LEOPARD contains 508 prostatectomy WSIs with biochemical recurrence annotations, while Miro-120 consists of 120 prostate H\&E-stained tissue samples annotated at the functional tissue unit level, comprising 18 radical prostatectomy specimens from the OHSU Biolibrary and 102 needle biopsy specimens from the CEDAR Biorepository.

\input{tables/patient_demographic}

\subsubsection{Data split}
A total of 819 patients were initially considered for inclusion from Radboud University Medical Center and Canisius Wilhelmina Hospital. Because the aim of CHIMERA was to provide a fully paired multimodal dataset, each case was required to contain patient characteristics, clinician-derived variables, digitized pathology images, and preoperative multiparametric MRI. The modality-specific inclusion criteria are described below. Cases were excluded when one or more required modalities were incomplete or unavailable, for example when the mpMRI, or specific MRI sequences were not available. In addition, data originated from multiple clinical departments and source systems, including radiology, pathology, and clinical records. Because all data were processed in anonymized form, cross-modality linkage could not always be completed with sufficient certainty. This contributed to the exclusion of incomplete or non-linkable cases. In total, 552 of 819 initially considered patients were excluded, resulting in 267 fully paired cases.

\begin{figure}[!ht]
    \centering
    \includegraphics[width=1\linewidth]{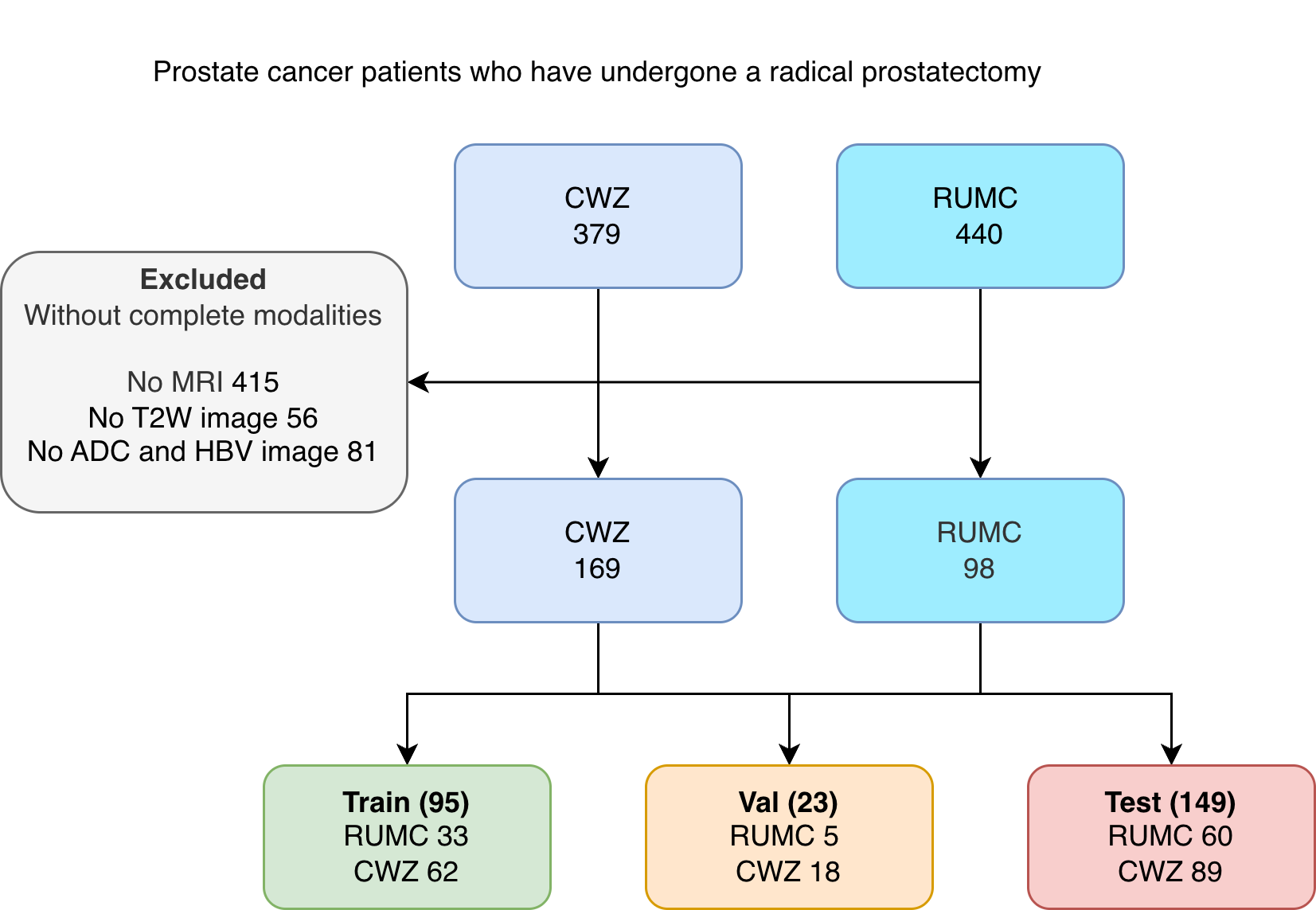}
    \caption{Data distribution of prostate cancer patients who have had a radical prostatectomy between 2012 and 2021 for Task 1 of the CHIMERA challenge.}
    \label{fig:data_distribution}
\end{figure}

Follow-up data and patient characteristics were extracted from anonymized pathology reports, which also formed the basis for designing the training, validation, and test splits. Because the CHIMERA dataset partially overlapped with the PI-CAI dataset \cite{Saha2024}, the data splits were designed to prevent leakage between publicly available PI-CAI data and the hidden CHIMERA evaluation sets. Cases included in the public PI-CAI set were only considered for the CHIMERA training set, whereas cases included in hidden PI-CAI sets were only considered for hidden CHIMERA splits. After applying these criteria, 95 cases were allocated to the training split, 23 to the validation split, and 149 to the test split. The resulting proportions were therefore determined by the PI-CAI overlap constraint rather than chosen a priori. An overview of the data distribution and excluded cases is provided in Fig. \ref{fig:data_distribution}. 

To assess consistency of data distributions across the three splits, statistical comparisons were performed. Continuous variables were analyzed using the Kruskal--Wallis test, while categorical variables were evaluated using the chi-square test for independence. The significance threshold was set at $\alpha = 0.05$.

\subsubsection{Pathology images}
WSIs were derived from the resected radical prostatectomy specimen and, depending on where the prostatectomy was performed, were prepared at the respective hospital-based pathology laboratories according to local tissue processing and H\&E staining protocols . All slides were subsequently digitized at RUMC using a 3DHISTECH scanner at a resolution of 0.25 $\mu$m/pixel. From an initial pool of 1,993 WSIs, experienced pathologists (19, 13, and eight years) reviewed all slides and selected the most prognostically informative ones, specifically those corresponding to the highest ISUP grade reported in the original pathology reports, yielding a final set of 942 WSIs.

All submitted algorithms were evaluated on the Grand Challenge platform using an NVIDIA T4 GPU with 16 GB of available VRAM. To limit computational overhead, the input format was restricted to a maximum of 10 WSIs per patient and a maximum file size of 10 GB per .tiff file. Differences in the number of WSIs per case were handled using a custom backend developed by the Grand Challenge team. As most patients had multiple WSIs, slides were grouped by merging up to two WSIs into each .tiff file and downsampling them to a resolution of 0.5 $\mu$m/pixel using an in-house developed algorithm. 

\subsubsection{Radiology images}
Preoperative mpMRI scans covering the prostate and surrounding pelvic region were acquired at the RUMC using 1.5T or 3.0T MRI scanners (Siemens Healthineers, Erlangen, Germany). The imaging protocol comprised three axial sequences: T2-weighted (T2W) imaging, high b-value diffusion-weighted imaging (HBV), and apparent diffusion coefficient (ADC) maps (Fig. \ref{fig:data overview}), yielding 801 sequences across 267 patients. Acquisition standards are consistent with those of the publicly released PI-CAI dataset \cite{Saha2024}.

\subsubsection{Patient characteristics \& clinician-derived variables}
In addition to imaging data, patient characteristics and clinician-derived variables were collected. Age and prostate-specific antigen (PSA) level at the time of surgery were recorded as continuous variables, and earlier therapy was recorded as a categorical variable indicating any treatment received before radical prostatectomy. The primary outcome measure was defined as the time to either BCR or last available follow-up, expressed as a continuous time-to-event variable in number of months. BCR was defined as a confirmed rising PSA following radical prostatectomy, verified per case against the treating institution's clinical documentation of biochemical recurrence. The lowest qualifying PSA value in the cohort was 0.1 ng/mL. A binary BCR status indicator was included to specify whether the event represented recurrence or censoring. For patients who experienced BCR, the PSA level at the time of recurrence was also recorded as an additional variable. However, this value was masked during model inference, as it directly reflects the occurrence of the event. Patients who did not experience BCR by the end of the five-year follow-up, or who were lost to follow-up, were treated as censored observations. As some variables were manually abstracted, human error cannot be excluded.

Clinician-derived variables encompassed pathological features routinely reported after radical prostatectomy, including the primary, secondary, and tertiary Gleason grades (ranging from 3 to 5), all recorded as integer values. The ISUP grade group was stored as an integer ranging from 1 to 5. Pathological tumor stage (pT stage) was recorded as stated in the histopathological report, alongside surgical margin status, capsular penetration, seminal vesicle invasion, lymph node involvement, and lymphovascular invasion. These variables were selected because they are established clinicopathological predictors of BCR \cite{EAUGuidelines2025}, and we hypothesized that including them would enable participants to identify novel prognostic signals in the pathology data.

In total, the released dataset contained 13 variables per patient (3 patient characteristics and 10 clinician-derived pathology variables) paired with a time-to-event ground truth label. All patient characteristics and clinician-derived data were stored in JSON format (Fig. \ref{fig:data overview}). For the validation and test phases of the challenge, BCR status, time to BCR or follow-up, and PSA levels at the time of recurrence were masked.

\subsubsection{Additional data}
\begin{figure}[H]
    \centering
    \includegraphics[width=1\linewidth]{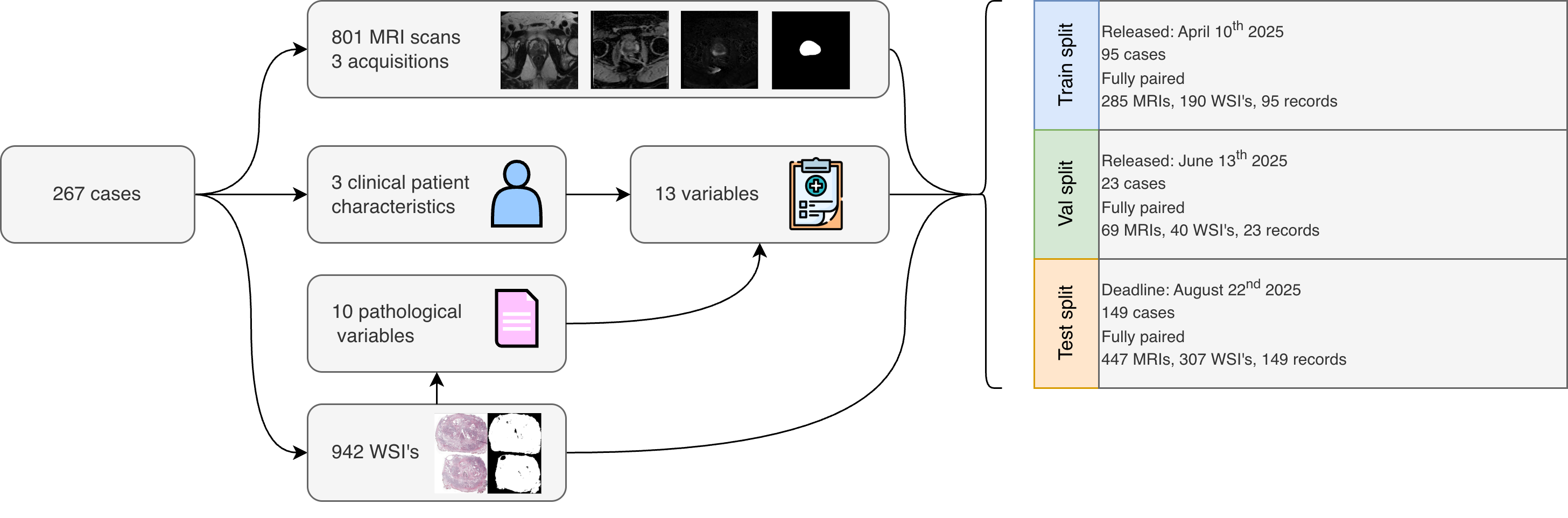}
    \caption{Flowchart of available data types provided in CHIMERA challenge.}
    \label{fig:data overview}
\end{figure}

In addition to the imaging data and variables that were collected, binary masks were generated for each individual WSI using the tissue segmentation algorithm previously developed by Bándi et al. \cite{Bandi2019}. These masks were packed analogously to the WSIs and provided alongside each corresponding .tiff file (Fig. \ref{fig:data overview}). 

Participants were free to use any feature encoder they preferred during the challenge. Nevertheless, to ensure sustainability and enable participants to develop deep learning methods without requiring substantial computational resources, feature embeddings were pre-extracted using UNI v1 \cite{Chen2024}. Each WSI was divided into patches of 224×224 pixels, from which 1024-dimensional feature embeddings were generated. The slide2vec package \cite{Slide2vec2023} was used to facilitate this extraction process.

For the mpMRI-images, a binary prostate gland segmentation mask was provided using the PI-CAI model from Saha et al. (Fig. \ref{fig:data overview}) \cite{PICAI_Study_design}. 

\subsection{Challenge Design and Organization}
The CHIMERA challenge consisted of three tasks. A full list of CHIMERA challenge organizers across all three tasks is provided in Supplementary Section \ref{sec:supp-organizers}. In this work we focus on task 1, a prognostic time-to-event prediction task aimed at predicting the time to BCR for prostate cancer patients after a radical prostatectomy, with the intended field of application being post-operative prognostic risk assessment. The algorithm target is the patient as a whole: algorithms produced a single patient-level prediction of the time to BCR, expressed in months, rather than localizing or delineating a specific anatomical structure. CHIMERA was organized as part of MICCAI 2025, where results were presented at a dedicated CHIMERA workshop. The design and reporting of this challenge follow the Biomedical Image Analysis Challenges (BIAS) reporting guidelines \cite{maierhein2020bias}, and the completed BIAS checklist is provided as supplementary material. The challenge was run as a one-time benchmark with a fixed submission deadline: although future editions with expanded cohorts and additional tasks are envisioned, the present challenge is not a continuous or repeated event, and the leaderboard reported here reflects a single submission cycle. The challenge was hosted on the Grand Challenge platform \cite{Meakin2025} in three phases: training (April 10, 2025), validation (June 13, 2025), and testing (August 22, 2025), and is accessible at \url{https://chimera.grand-challenge.org/chimera/}. Team registration opened on May 27, 2025. For each case across all three splits, participants had access to all available modalities, comprising mpMRI, WSI, and patient characteristics including clinician-derived variables from the Radboud University Medical Center and the Canisius Wilhelmina Hospital, both in Nijmegen, the Netherlands. Participants were not required to use all modalities and were free to design their algorithms using any subset of the available data. Only fully automatic algorithms were permitted. Participants were encouraged to make the source code of their submitted models publicly available.

The training data was hosted on the Amazon Web Services (AWS) platform. In the validation phase, participating teams were allowed to submit their algorithms up to five times to the Grand Challenge platform as Docker container images to receive performance metrics on the validation set. During the testing phase, participants were permitted to submit their final algorithm only once, which was then evaluated on the test set. An example baseline codebase demonstrating the required submission interface, together with the evaluation code used to compute the challenge metrics, was provided at \url{https://github.com/nadieh/CHIMERA_minimal_baseline}. The property to be optimized was the ability to correctly order patients by their time to BCR, rather than to estimate event times accurately in absolute terms. Final submissions were ranked based on the concordance index (C-index) \cite{Harrell1982}, as it assesses risk ranking while appropriately accounting for right-censored observations. This allows for a fair, model-agnostic comparison of survival models under the heterogeneous follow-up conditions present in the dataset. The C-index was computed once over the full test set and teams were ranked in descending order of this single aggregated value.

Members of the organizing institutes were permitted to participate in the challenge but were not eligible for awards; in practice, no team affiliated with an organizing institute submitted an algorithm. Cash prizes were awarded per task based on the final test-set ranking: €1000 for first place, €600 for second, and €400 for third. Participants affiliated with organizing centers or the sponsoring company were not eligible for cash prizes. The final rankings, based on the test-set C-index, were announced publicly during the CHIMERA workshop on September 23, 2025. For each task, the top-ranked teams were invited to join the CHIMERA Consortium, with up to five co-authors per team included as consortium authors on this paper. Participants were permitted to publish their own results, subject to an embargo until publication of the CHIMERA challenge paper, and were required to cite the challenge publication.

Reference labels for the validation and test sets, comprising BCR status, time to BCR or last available follow-up, and PSA level at recurrence, were held exclusively by the organizing team at RUMC and were never released to participants at any phase of the challenge. All evaluations on the hidden validation and test sets, including the post-hoc analyses reported here, were performed by the organizers on containerized algorithms supplied by the participating teams.

The present study was approved by the Institutional Review Board of RUMC (2022–15,878). Informed consent was waived due to the usage of anonymized prostate specimens and radiological scans in a retrospective setting. The challenge data is released under the Creative Commons Attribution-Non\-Commercial-Share\-Alike (CC BY-NC-SA) license, permitting non-commercial use and redistribution provided that appropriate attribution is given and derivative datasets are shared under the same terms. Use of the data is additionally subject to the publication embargo described above.

\subsection{Analysis}
First, we summarized the performance of each submitted model during the challenge. We also collected and reviewed the methodological information for each model, including the modalities used, preprocessing steps, feature extraction strategy, and fusion approach, to characterize design choices across submissions.

Some submitted models were trained exclusively on patient characteristics and clinician-derived variables. To assess the contribution of individual pathological variables to model performance, we performed a post-hoc feature importance analysis using randomization at test time. Because several models relied on lookup tables or pre-processing pipelines that expected valid clinical values, setting variables to zero or omitting them entirely would either cause errors in model execution or, for binary variables where zero encodes a clinically meaningful absence, introduce systematic bias rather than a neutral perturbation. Instead, each variable was replaced with a synthetic value drawn from its valid clinical range: binary variables were sampled uniformly from \{0,1\}, and categorical and ordinal variables were sampled uniformly from their respective valid ranges. Variables were randomized individually, in semantically related groups (for example, primary, secondary, and tertiary Gleason grade randomized jointly), and as a complete clinician-derived subset.

Clinician-derived variables include pathology-derived measures such as Gleason grade and ISUP grade group, which in principle could also be learned directly from pathology images. Providing these variables in tabular form may therefore act as a shortcut that masks the true contribution of imaging data to multimodal model performance. To evaluate this, multimodal models were retrained on a modified version of the dataset with all clinician-derived pathological variables removed. This tests whether multimodal models can retain predictive performance by leveraging the imaging data alone, without access to the tabular shortcut.

For one multimodal model, we additionally conducted a dedicated modality ablation analysis by retraining across all combinations of MRI, WSI, Clin\_full, and Clin\_filtered, where Clin\_full contains both patient characteristics and clinician-derived variables and Clin\_filtered denotes the dataset after removal of clinician-derived variables. As retraining across all modality combinations represented a substantial additional burden on participants, this model was selected based on team availability to accommodate the additional retraining requests within the study timeline. Within this single multimodal framework, the ablation provides a more granular view of the relative and combined contributions of each modality, allowing us to assess whether modalities provide complementary prognostic information or whether performance is dominated by a single input type.

Statistical significance was assessed by bootstrapping model performance over 10,000 iterations, resampling test cases with replacement. Bootstrapping was chosen over parametric alternatives because the concordance index has no convenient closed-form variance under censoring and the comparisons are paired within patients. For the unimodal feature importance analysis and the modality importance analysis for one multimodal model, two sided p-values were adjusted for multiple testing using Benjamini-Hochberg false discovery rate correction (BH-FDR) and reported as $q$-values. BH-FDR was preferred over family-wise error rate control given the exploratory nature of these post-hoc analyses and the correlation between nested randomization conditions. A significance threshold of $\alpha = 0.05$ was applied throughout. All statistical analyses were performed in Python (3.11) using scikit-survival, SciPy, and NumPy. For the post-hoc experiments, Docker containers were provided by participating teams and were not part of the original competition submissions.

%% file: tables/patient_demographic.tex
\begin{table}[p]
    \thispagestyle{empty} 
    \vspace*{-2.5cm} 
    \centering
    \caption{Patient characteristics and clinical features stratified by data split. Continuous variables are presented as median (IQR); categorical as n (\%). P-values were calculated using the Kruskal-Wallis test (continuous) and Pearson's Chi-squared test (categorical).}
    \vspace{0.2cm}
    \label{tab:patient_demographic}
    \scriptsize 
    \renewcommand{\arraystretch}{0.80} 
    
    \makebox[\textwidth][c]{
        \begin{threeparttable}
            \begin{tabular}{l c c c c}   
                \hline
                \textbf{Variable} & \textbf{Train (n=95)} & \textbf{Val (n=23)} & \textbf{Test (n=149)} & \textbf{\textit{p}-value} \\
                \hline
                \textbf{Age}, \textit{median (IQR)} & 66.0 (60.0 - 69.0) & 66.0 (62.0 - 68.0) & 65.0 (60.0 - 68.0) & 0.693 \\
                \textbf{PSA}, \textit{median (IQR)} & 7.8 (5.2 - 12.0) & 10.7 (6.2 - 17.0) & 8.4 (5.8 - 12.7) & 0.244 \\
                
                \textbf{BCR}, \textit{n (\%)} &  &  &  & 0.953 \\
                \hspace{3mm} 0 & 68 (71.6\%) & 16 (69.6\%) & 104 (69.8\%) &  \\
                \hspace{3mm} 1 & 27 (28.4\%) & 7 (30.4\%) & 45 (30.2\%) &  \\
                
                \textbf{Lymphovascular Invasion}, \textit{n (\%)} &  &  &  & 0.615 \\
                \hspace{3mm} 0 & 78 (82.1\%) & 19 (82.6\%) & 129 (86.6\%) &  \\
                \hspace{3mm} 1 & 17 (17.9\%) & 4 (17.4\%) & 20 (13.4\%) &  \\
                
                \textbf{Primary Gleason}, \textit{n (\%)} &  &  &  & 0.812 \\
                \hspace{3mm} 2 & 0 (0.0\%) & 0 (0.0\%) & 2 (1.3\%) &  \\
                \hspace{3mm} 3 & 54 (56.8\%) & 12 (52.2\%) & 88 (59.1\%) &  \\
                \hspace{3mm} 4 & 39 (41.1\%) & 10 (43.5\%) & 57 (38.3\%) &  \\
                \hspace{3mm} 5 & 2 (2.1\%) & 1 (4.3\%) & 2 (1.3\%) &  \\
                
                \textbf{Secondary Gleason}, \textit{n (\%)} &  &  &  & 0.901 \\
                \hspace{3mm} 2 & 1 (1.1\%) & 0 (0.0\%) & 4 (2.7\%) &  \\
                \hspace{3mm} 3 & 36 (37.9\%) & 11 (47.8\%) & 55 (36.9\%) &  \\
                \hspace{3mm} 4 & 48 (50.5\%) & 10 (43.5\%) & 74 (49.7\%) &  \\
                \hspace{3mm} 5 & 10 (10.5\%) & 2 (8.7\%) & 16 (10.7\%) &  \\
                
                \textbf{Tertiary Gleason}, \textit{n (\%)} &  &  &  & 0.872 \\
                \hspace{3mm} 3 & 2 (2.1\%) & 1 (4.3\%) & 6 (4.0\%) &  \\
                \hspace{3mm} 4 & 3 (3.2\%) & 1 (4.3\%) & 4 (2.7\%) &  \\
                \hspace{3mm} 5 & 15 (15.8\%) & 2 (8.7\%) & 16 (10.7\%) &  \\
                \hspace{3mm} x\tnote{a} & 75 (78.9\%) & 19 (82.6\%) & 123 (82.6\%) &  \\
                
                \textbf{ISUP}, \textit{n (\%)} &  &  &  & 0.870 \\
                \hspace{3mm} 1 & 8 (8.4\%) & 4 (17.4\%) & 22 (14.8\%) &  \\
                \hspace{3mm} 2 & 44 (46.3\%) & 8 (34.8\%) & 59 (39.6\%) &  \\
                \hspace{3mm} 3 & 27 (28.4\%) & 7 (30.4\%) & 42 (28.2\%) &  \\
                \hspace{3mm} 4 & 7 (7.4\%) & 1 (4.3\%) & 12 (8.1\%) &  \\
                \hspace{3mm} 5 & 9 (9.5\%) & 3 (13.0\%) & 14 (9.4\%) &  \\
                
                \textbf{pT Stage}, \textit{n (\%)} &  &  &  & 0.126 \\
                \hspace{3mm} 2 & 51 (53.7\%) & 13 (56.5\%) & 79 (53.0\%) &  \\
                \hspace{3mm} 3a & 28 (29.5\%) & 10 (43.5\%) & 51 (34.2\%) &  \\
                \hspace{3mm} 3b & 11 (11.6\%) & 0 (0.0\%) & 18 (12.1\%) &  \\
                \hspace{3mm} 4 & 5 (5.3\%) & 0 (0.0\%) & 1 (0.7\%) &  \\
                
                \textbf{Positive Lymph Nodes}, \textit{n (\%)} &  &  &  & 0.853 \\
                \hspace{3mm} 0 & 32 (33.7\%) & 9 (39.1\%) & 56 (37.6\%) &  \\
                \hspace{3mm} 1 & 10 (10.5\%) & 1 (4.3\%) & 16 (10.7\%) &  \\
                \hspace{3mm} x\tnote{b} & 53 (55.8\%) & 13 (56.5\%) & 77 (51.7\%) &  \\
                
                \textbf{Capsular Penetration}, \textit{n (\%)} &  &  &  & 0.350 \\
                \hspace{3mm} 0 & 52 (54.7\%) & 13 (56.5\%) & 84 (56.4\%) &  \\
                \hspace{3mm} 1 & 39 (41.1\%) & 10 (43.5\%) & 64 (43.0\%) &  \\
                \hspace{3mm} x\tnote{b} & 4 (4.2\%) & 0 (0.0\%) & 1 (0.7\%) &  \\
                
                \textbf{Positive Surgical Margins}, \textit{n (\%)} &  &  &  & 0.307 \\
                \hspace{3mm} 0 & 45 (47.4\%) & 12 (52.2\%) & 84 (56.4\%) &  \\
                \hspace{3mm} 1 & 50 (52.6\%) & 11 (47.8\%) & 62 (41.6\%) &  \\
                \hspace{3mm} x\tnote{b} & 0 (0.0\%) & 0 (0.0\%) & 3 (2.0\%) &  \\
                
                \textbf{Seminal Vesicle Invasion}, \textit{n (\%)}&  &  &  & 0.317 \\
                \hspace{3mm} 0 & 81 (85.3\%) & 23 (100.0\%) & 131 (87.9\%) &  \\
                \hspace{3mm} 1 & 14 (14.7\%) & 0 (0.0\%) & 17 (11.4\%) &  \\
                \hspace{3mm} x\tnote{b} & 0 (0.0\%) & 0 (0.0\%) & 1 (0.7\%) &  \\
                
                \textbf{Earlier Therapy}, \textit{n (\%)} &  &  &  & 0.450 \\
                \hspace{3mm} None & 92 (96.8\%) & 23 (100.0\%) & 146 (98.0\%) &  \\
                \hspace{3mm} Other & 0 (0.0\%) & 0 (0.0\%) & 3 (2.0\%) &  \\
                \hspace{3mm} Radiotherapy + cryotherapy & 1 (1.1\%) & 0 (0.0\%) & 0 (0.0\%) &  \\
                \hspace{3mm} Radiotherapy + hormones & 1 (1.1\%) & 0 (0.0\%) & 0 (0.0\%) &  \\
                \hspace{3mm} Unknown & 1 (1.1\%) & 0 (0.0\%) & 0 (0.0\%) &  \\
                \hline
            \end{tabular} 
            
            \begin{tablenotes}
                \footnotesize
                \item[a] x: Indicates the feature was not present in the patient (e.g., no tertiary Gleason pattern).
                \item[b] x: Indicates data was not available or not extracted (e.g., lymph nodes were not removed during surgery).
            \end{tablenotes}
            
        \end{threeparttable}
    } 
\end{table}

%% file: sections/results.tex
\section{Results}
\label{sec:results}

\subsection{Challenge Leaderboard and Overview of Evaluated Algorithms}

\begin{figure}[!ht]
    \centering
    \includegraphics[width=1\linewidth]{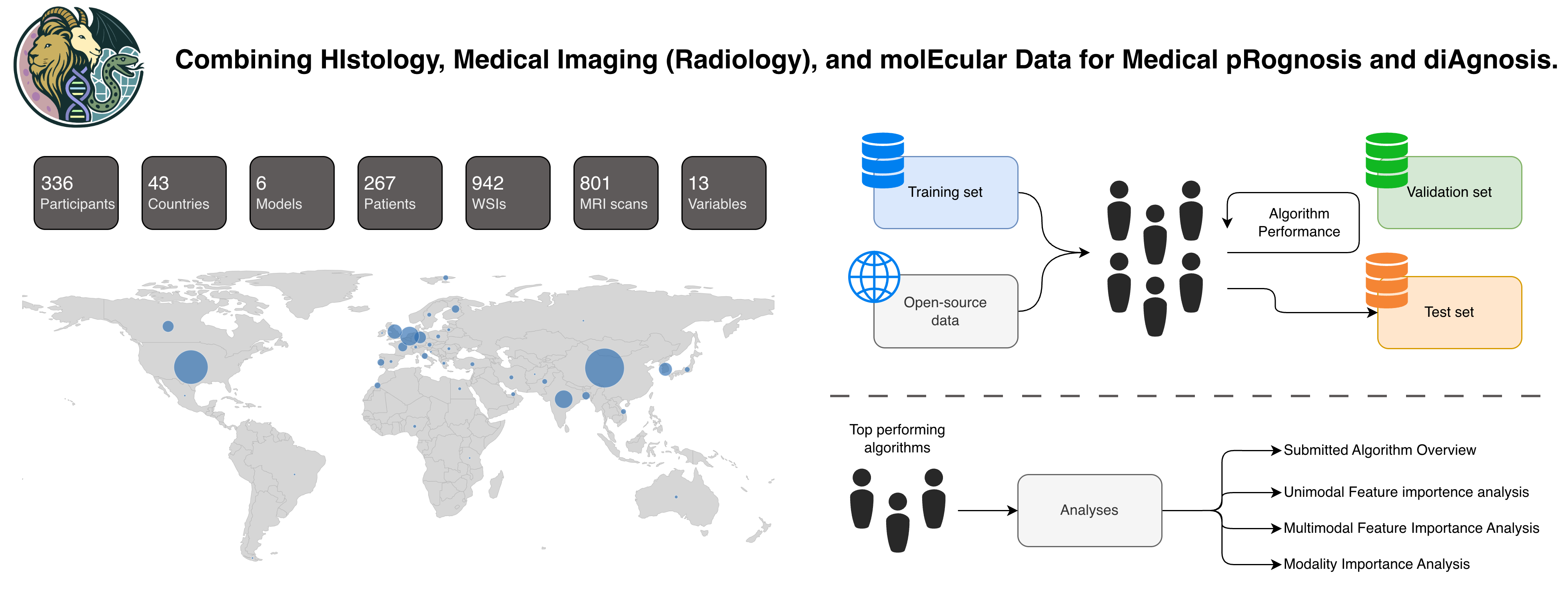}
    \caption{\textbf{Overview of the CHIMERA challenge and study setup.} The global competition attracted 336 participants from 43 countries. The challenge was organized into three phases. Participants could download the open-source training set for algorithm development and submit candidate algorithms for evaluation on the hidden validation and test sets through the Grand Challenge platform. Each participant was allowed up to five successful submissions to the validation set and one successful submission to the final test set. After the challenge ended, algorithms that were successfully submitted to the test phase were included in additional analyses, including feature importance and modality importance analyses.}
    \label{fig:CHIMERA_overview_setup}
\end{figure}

During the CHIMERA challenge, 336 participants from 43 countries took part. 21 teams were officially registered and six submitted algorithms for evaluation. The largest participant groups came from China, the United States, India, the Netherlands, and the United Kingdom. Throughout the challenge, participants could submit algorithm versions for evaluation during the development and debug phases before the final test phase, of which six teams submitted a final model in the testing phase (Fig. \ref{fig:CHIMERA_overview_setup}).

For each of the six teams, we summarize their methodology in terms of data handling and preprocessing, feature encoding, fusion strategy (if applicable), and model performance. Final leaderboard results for all participating teams are presented in Table \ref{tab:leaderboard_results}. The top three models employed only patient characteristics and clinician-derived variables. One team combined the patient characteristics and clinician-derived variables with the WSIs. Two teams submitted full multimodal models utilizing all available data for each case.

\input{tables/overview_evaluated_algorithms}

\subsubsection{TIA-Pegasus}
The TIA-Pegasus team introduced ModalSurv, integrating patient characteristics, clinician-derived variables, mpMRI, and WSIs through modality-specific projection heads and a cross-attention fusion mechanism. Pathology representations were derived from CONCH tile embeddings aggregated into 768-dimensional WSI-level features using the Titan foundation model for a random selection of 500 patches per WSI. Radiology features were extracted from all three provided mpMRI acquisitions (T2W, HBV, and ADC) using a pretrained MedicalNet ResNet-50 encoder after which global average pooling produced 2048-dimensional radiology embeddings. Clinical variables were cleaned by excluding entries with missing values, then encoded as normalized tabular features, resulting in a 10-dimensional feature vector. Each modality was projected into a shared latent space and fused using cross-attention, and the resulting representation was optimized within a discrete-time DeepHit survival framework.

Although ModalSurv was designed as a fully multimodal architecture, the team ultimately submitted a "clinical-only" model (using only patient characteristics and clinician-derived variables). During internal evaluation, their clinical and WSI combined model performed best (C-index of 0.8770). However, on the CHIMERA validation set, performance of that configuration dropped substantially to 0.6446, suggesting limited generalization and possible overfitting given the small cohort size. An additional source of variability is ModalSurv's use of 500 randomly sampled patches per WSI, which, combined with domain differences between the training and validation data, may have further degraded performance.

On the validation leaderboard, the TIA-Pegasus models performed best using only patient characteristics and clinician-derived variables, achieving a C-index of 0.8182. This clinical-only model subsequently attained the top test-phase performance with a C-index of 0.7402.

\subsubsection{WL}
The WL team submitted a clinical-only model using patient characteristics and clinician-derived variables. During preprocessing, variables were encoded either as ordinal variables or using one-hot encoding. Follow-up times were discretized into fixed intervals based on the distribution of event times in the training data, and for censored samples the time was set to the maximum observed interval.

To account for the small cohort size and improve generalizability across training and validation splits, the WL team used 20 cross-validation splits encompassing 10-fold cross-validation under two stratification criteria: age and time-to-follow-up or BCR. Six model types were evaluated: Cox Proportional Hazards, Random Survival Forest, Survival XGBoost, and three neural network-based models including MLP, RealMLP, and TabM. The neural models were formulated within a discrete-time survival analysis framework and trained using a negative log-likelihood loss.

For each split, all six models were trained independently. Based on performance on the internal validation split, the top three models were selected for ensemble, with final predictions computed by averaging outputs across all splits and model combinations, yielding an average over 60 outputs to reduce variance. However, on the CHIMERA validation set, a standard MLP model outperformed the ensemble, achieving a C-index of 0.7521 compared to 0.7423.

The WL team also performed an ablation study comparing fine-grained time discretization using 110 bins against coarse-grained discretization using 11 bins. Across all three top-performing models, the coarser binning yielded better performance on the internal validation split. The standard MLP model with 11 time bins was therefore selected for the final test phase, achieving a C-index of 0.7294.

\subsubsection{SMILE}
The SMILE team submitted a clinical-only model using patient characteristics and clinician-derived variables. During preprocessing, missing values were imputed throughout the dataset. For the tertiary Gleason score specifically, all missing values were set to 2, as all other observed values for this variable in the dataset were consistently 3 or higher. For all other variables, missing values were imputed using the mode.

Variables were categorized into three types: numeric, ordinal, and categorical. Each type was preprocessed accordingly through imputation, normalization, and one-hot encoding, respectively. For categorical variables, missing values were filled with a sentinel token "missing" prior to encoding. One-hot encoding expanded each categorical variable into mutually exclusive binary indicators, with unseen categories ignored during inference. The outputs of these preprocessing steps were horizontally concatenated and passed into a three-layer MLP producing four logits, each corresponding to one of four time bins. Hazard probabilities for each time bin were obtained by applying a sigmoid transformation to the logits, and the final survival output was derived as the cumulative product of the predicted hazard probabilities.

The model was trained using a negative log-likelihood loss with 3-fold cross-validation, using an 80/20 training and validation split per fold. Patients were stratified by BCR status and across discrete risk bins derived from the survival time distribution. This approach achieved a C-index of 0.7355 on the validation set and 0.7280 on the test set.

\subsubsection{IU CompPath}
The IU CompPath team introduced PROFUSEme, integrating patient characteristics and clinician-derived variables, mpMRI, and WSIs through intermediate fusion of modality-specific embeddings \cite{you2025profuseme}. Of the available patient characteristics and clinician-derived variables, eight were selected: age at radical prostatectomy, ISUP grade, pT stage, positive lymph nodes, capsular penetration, positive surgical margins, seminal vesicle invasion, and lymphovascular invasion. Categorical variables were one-hot encoded and numerical variables were Z-score normalized, resulting in a 25-dimensional clinical vector. Radiology features were extracted from co-registered HBV, ADC, and T2W MRI sequences using MRI-CORE \cite{dong2025mricore}. Slice-wise MRI embeddings were flattened, reduced using fully connected layers, and aggregated with self-attention pooling to produce 512-dimensional patient-level radiology vectors.

In addition to the CHIMERA training set, the IU CompPath team also employed the LEOPARD challenge dataset \cite{grisi2026leopard, faryna2026leopard}, contributing 508 additional WSI cases to pathology model development. These additional WSIs were used in the two-stage pathology pipeline rather than as part of the paired multimodal CHIMERA cohort. Pathology features were extracted from WSIs using a previously developed two-stage “thinking fast" and "thinking slow” pipeline. In the “thinking fast” stage, recurrence was modeled as an auxiliary binary classification task using a 22-month threshold, excluding cases without BCR and with follow-up shorter than 22 months. In the “thinking slow” stage, the model was trained for time-to-recurrence prediction using all CHIMERA training WSIs and all LEOPARD training WSIs. WSIs were segmented using CLAM, color-deconvolved to H-E-DAB space, filtered for foreground tissue, and patched at low and high resolution. Patch embeddings were generated using pathology foundation models, with the final selected configuration using UNI V2 for the lower-resolution stage and UNI V1 for the higher-resolution stage. Patch-level embeddings were reduced, selected using top-k feature selection, and aggregated with self-attention pooling to produce 512-dimensional patient-level pathology vectors.

The model projected the clinical, radiology, and pathology vectors into a shared 768-dimensional token space and stacked them as modality tokens. A four-layer Transformer encoder with multi-head self-attention was then used to model cross-modal interactions, followed by mask-aware mean pooling and a linear Cox proportional hazards survival head that predicted log-risk (LR). Mask-aware mean pooling was applied to handle missing modalities if present. The final time-to-recurrence prediction was obtained as $\exp(-LR)$, with the submitted model averaging log-risk predictions across nine cross-validation folds.

In internal 5-fold nested cross-validation, the model achieved a C-index of 0.8610. However, on the CHIMERA leaderboards, this approach achieved a C-index of 0.7107 on the validation set and 0.7197 on the test set, being the only model with a higher C-index on the test split compared to the validation split.

\subsubsection{VCMI}
The VCMI team introduced a multimodal discrete-time survival model, integrating patient characteristics and clinician-derived variables, mpMRI, and WSIs through modality-specific branches followed by gated late fusion. Clinical data were encoded using both manually engineered tabular features and text-based embeddings. The manually engineered representation consisted of 21 features, including missingness masks for selected variables. In parallel, a structured text prompt containing the clinical variables was embedded using a Sentence Transformer model. These tabular and text-derived representations were concatenated and processed by a residual MLP to produce a 128-dimensional clinical embedding. 

Radiology features were extracted from T2W, ADC, and HBV MRI sequences after N4 bias field correction, resampling to 1 × 1 × 1 mm isotropic spacing, registration to the T2W reference, and z-score normalization. Each MRI sequence was encoded using a Med3D-pretrained MONAI ResNet101, producing 2048-dimensional sequence-level features that were projected to 256-dimensional embeddings and combined using learnable weights. 

Pathology features were extracted from WSIs by removing non-informative background regions using the provided masks, extracting 512 × 512 patches with TRIDENT, and embedding patches with UNI V1. Patch embeddings were aggregated into slide-level representations using gated ABMIL, and patient-level pathology embeddings were obtained by averaging across all available slides, resulting in a 512-dimensional pathology representation.

Each modality embedding was projected and assigned a learned sample-specific weight through gated late fusion. The weighted embeddings were concatenated into an 896-dimensional multimodal vector and processed by an MLP to output probabilities across 15 discrete time bins. Survival modeling was performed using DeepHit. The final submission used a modified Weibull binning strategy, in which bin boundaries were primarily derived from a Weibull distribution fitted to uncensored recurrence times. To improve granularity among short time-to-recurrence cases, early time ranges were subsequently subdivided using uniform binning. Predicted logits were converted into interval-specific hazard probabilities, survival probabilities were computed cumulatively across time bins, and the final risk score was derived from the summed negative survival probabilities.

For the team's final model, they first used stratified 75/25 splits for model selection and hyperparameter tuning. After selecting the best configuration, they trained the final model on the full dataset achieving a training C-index of 0.97. On the CHIMERA leaderboards the model achieved a C-index of 0.7521 for the validation set and a C-index of 0.7153 for the final test set. 

\subsubsection{OHSU-Cedar}
The OHSU-Cedar team submitted a model incorporating patient characteristics, clinician-derived variables, and WSIs through a biologically informed Functional Tissue Unit (FTU) classification pipeline. Patient characteristics and clinician-derived variables were preprocessed according to data type. Ordinal variables were assigned integer ranks, resulting in a 13-dimensional clinical feature vector. Missing values were imputed using the training-set mean for continuous variables and the training-set mode for binary variables.

The pathology branch used an FTU classification model based on a Virchow2 backbone, trained on the Miro-120 dataset \cite{Fenner2025}, to categorize tissue tiles into six biologically meaningful prostate tissue classes corresponding to benign tissue and Gleason-related patterns. For each CHIMERA WSI, the trained FTU classifier was applied to 2000 randomly sampled tiles, after which the relative proportions of the six FTU categories were aggregated into a 6-dimensional histogram-based pathology feature vector.

The 13-dimensional clinical vector and 6-dimensional pathology vector were concatenated into a 19-dimensional feature representation, followed by z-score normalization using statistics computed on the training set. During inference, the same training-set mean and standard deviation were used for normalization. For patients with multiple WSIs, only the first WSI was used. A 2-layer MLP with ReLU activation after the first layer and a sigmoid output layer was trained on these embeddings using a hinge loss as training objective. Cross-validation was performed using repeated random splits with 70\% of the data used for training. On the CHIMERA leaderboards, the approach achieved a C-index of 0.7686 on the validation set and 0.6885 on the test set.

\subsection{Unimodal Feature Importance Analysis}
For unimodal models, we report the contribution of clinician-derived clinical variables to predictive performance by randomizing these variables. Subsequently, the challenge models employing only patient characteristics and clinician-derived variables were re-run on these altered datasets. Table \ref{tab:feature_importance_UN} reports the bootstrapped $\Delta$C-index ($C_0 - C_r$) with 95\% confidence intervals and BH-FDR adjusted q-values. Across models, positive lymph node status and Gleason grading were the most consistently impactful variables. 

For the WL model ($C_0 = 0.7294$), randomizing positive lymph nodes resulted in the largest performance drop ($\Delta C = 0.1292$, $q = 0.012$), followed by primary Gleason ($\Delta C = 0.0889$, $q = 0.025$). Randomizing all pathology parameters simultaneously produced a $\Delta C$ of 0.1790 ($q = 0.033$). 

Similarly, for SMILE ($C_0 = 0.7280$), positive lymph nodes had the strongest individual impact ($\Delta C = 0.1017$, $q = 0.028$). Secondary Gleason ($\Delta C = 0.0397$, $q = 0.047$) and all Gleason components combined ($\Delta C = 0.0630$, $q = 0.046$) also significantly reduced performance. Simultaneous randomization of all pathology variables resulted in $\Delta C = 0.2112$ ($q = 0.028$).

For TIA-Pegasus ($C_0 = 0.7402$), primary Gleason ($\Delta C = 0.0642$, $q = 0.010$) and randomization of all pathology parameters showed the largest effects ($\Delta C = 0.2519$, $q = 0.001$). In contrast to WL and SMILE, randomizing positive lymph nodes had no impact on performance ($\Delta C = 0.0000$), suggesting differential feature reliance across models.

Several variables demonstrated minimal or inconsistent impact across all three models. Capsular penetration, tertiary Gleason, pT stage, pre-operative PSA, and lymphovascular invasion generally yielded small $\Delta C$ values with confidence intervals overlapping zero and non-significant $q$-values, suggesting limited independent contribution in the presence of other predictors.

When all pathology-derived parameters were randomized simultaneously, discriminative performance approached near-random levels ($C \approx$ 0.50–0.56), confirming that the predictive performance of unimodal clinical models is predominantly driven by structured pathological variables.

Results split per center can be found in Supplementary Figure \ref{sup_fig:forest_unimodal_feature_importance}

\input{tables/feature_importance_unimodal_models}

\subsection{Multimodal performance using no pathology-derived variables}
For multimodal models (models employing any form of imaging data in combination with patient characteristics and clinician-derived variables), we compare performance with and without clinician-derived variables, providing insight into whether imaging modalities can compensate for the absence of these variables, simulating clinical reality more closely. The outcomes of the retrained multimodal models are summarized in Table~\ref{tab:feature_importance_MM}. Clinician-derived variables were removed and models were retrained using only patient characteristics and imaging data. Performance differences are reported as $\Delta$C-index ($C_0 - C_r$), with BH-FDR-adjusted $q$-values.

For the IU-CompPath model, which excluded patient characteristics as well, relying solely on imaging data, performance decreased from $C_0 = 0.7197$ to $0.6869$, corresponding to $\Delta C = 0.0328$. The VCMI model performance declined from $C_0 = 0.7153$ to $0.6762$ ($\Delta C = 0.0391$). The OHSU-Cedar model demonstrated minimal sensitivity to the removal of pathology-derived variables, with performance decreasing from $C_0 = 0.6885$ to $0.6881$ ($\Delta C = 0.0004$).

No statistically significant performance reductions were observed across any of the three models (all $p > 0.05$), suggesting that predictive performance was largely preserved after removal of pathology-derived structured variables.

Results split per center can be found in Supplementary Figure \ref{sup_fig:forest_multimodal_feature_importance}

\input{tables/feature_importance_multimodal_models}

\subsection{Analyzing Modality Importance in the VCMI Model}
To characterize the contribution of individual modalities, we performed an ablation analysis of the VCMI model across combinations of WSI, MRI, patient characteristics, and clinician-derived variables, where Clin\_full contains both patient characteristics and clinician-derived variables and Clin\_filtered denotes the dataset after removal of clinician-derived variables. The full model, integrating WSI, MRI, and Clin\_full, was used as the baseline model configuration ($C_0 = 0.7153$). Results are summarized in Table \ref{tab:ablation_vcmi}.

\input{tables/modality_importance_VCMI}

Among the single-modality configurations, Clin\_full achieved the highest C-index and was not significantly different from the full multimodal reference model ($C_r = 0.6831$; $\Delta C = 0.0322$, $q = 0.4402$). WSI alone achieved moderate performance, but showed a significant decrease compared with the reference model ($C_r = 0.6460$; $\Delta C = 0.0693$, $q = 0.0355$). MRI alone performed lowest among all configurations ($C_r = 0.4595$; $\Delta C = 0.2558$, $q = 0.0040$).

Adding other modalities to WSI generally increased performance. WSI combined with Clin\_full achieved the closest performance to the reference model ($C_r = 0.7076$; $\Delta C = 0.0077$, $q = 0.5435$), while WSI combined with MRI also improved over WSI alone ($C_r = 0.6902$; $\Delta C = 0.0251$, $q = 0.4254$). In contrast, adding MRI to Clin\_full did not improve performance ($C_r = 0.6687$ compared with $C_r = 0.6831$ for Clin\_full alone). These results suggest that MRI contributed inconsistently across configurations: it improved performance when combined with WSI, but did not provide additional benefit when combined with the full set of patient characteristics and clinician-derived variables.

Configurations using only Clin\_filtered showed substantially lower performance. Clin\_filtered alone achieved near-random discrimination ($C_r = 0.5062$; $\Delta C = 0.2091$, $q = 0.0073$), and MRI combined with Clin\_filtered performed similarly ($C_r = 0.5010$; $\Delta C = 0.2143$, $q = 0.0073$). However, adding Clin\_filtered to WSI improved performance compared with WSI alone ($C_r = 0.6981$ versus $C_r = 0.6460$), suggesting that age and PSA provided complementary information in the presence of histopathological features. This improvement was not observed when Clin\_filtered was added to WSI and MRI ($C_r = 0.6762$ versus $C_r = 0.6902$), indicating that the contribution of Clin\_filtered was configuration-dependent. Overall, Clin\_full remained the most informative structured-data configuration, but Clin\_filtered was not uniformly uninformative across modality combinations.

Results split per center can be found in Supplementary Figure \ref{sup_fig:forest_modality_ablation}

%% file: tables/overview_evaluated_algorithms.tex
\begin{table}[h]
    \centering
    \caption{CHIMERA challenge leaderboard and overview of evaluated algorithms. Rows are sorted by final Test C-Index.}
    \vspace{0.2cm}
    \label{tab:leaderboard_results}
    
    \scriptsize 
    \renewcommand{\arraystretch}{1.2} 
    
    \makebox[\textwidth][c]{
        \begin{tabular}{c l c c c c}
            \hline
            \textbf{Rank} & \textbf{Team} & \textbf{Modalities} & \textbf{Val. C-Index} & \textbf{Test C-Index} & \textbf{Code} \\
            \hline
            1 & TIA-Pegasus & Clinical & 0.8182 & 0.7402 & Private \\
            2 & WL & Clinical & 0.7521 & 0.7294 & Private \\
            3 & SMILE & Clinical & 0.7355 & 0.7280 & \href{https://github.com/XulinChen/Algorithm-for-Chimera-Challenge}{GitHub} \\
            4 & IU CompPath & Clin + MRI + WSI & 0.7107 & 0.7197 & Private \\
            5 & VCMI & Clin + MRI + WSI & 0.7521 & 0.7153 & \href{https://github.com/fcoelhomrc/CHIMERA_VCMI}{GitHub} \\
            6 & OHSU-Cedar & Clin + WSI & 0.7686 & 0.6885 & Private \\
            \hline
            \multicolumn{6}{l}{\footnotesize \textit{Note:} ``Private'' indicates a repository that exists but is not publicly accessible at the} \\
            \multicolumn{6}{l}{\footnotesize  time of writing.} \\
        \end{tabular}
    }
\end{table}

%% file: tables/feature_importance_unimodal_models.tex
\begin{table}[H]
\centering
\caption{Overview of performance per unimodal team for randomized variables. Models were bootstrapped (10,000 samples). Baseline performance is denoted by $C_0$ and post-randomization performance by $C_r$; $\Delta$C-index ($C_0 - C_r$) is reported with 95\% bootstrap confidence intervals. Benjamini--Hochberg (BH) FDR-adjusted two-sided $p$-values ($q$) are shown separately. Values in bold denote $q < 0.05$.}
\vspace{0.2cm}
\label{tab:feature_importance_UN}

\scriptsize
\renewcommand{\arraystretch}{1.25}

\resizebox{\textwidth}{!}{
\begin{tabular}{l | ccc | ccc | ccc}
\hline
& \multicolumn{3}{c}{\textbf{WL}}
& \multicolumn{3}{c}{\textbf{SMILE}}
& \multicolumn{3}{c}{\textbf{TIA-Pegasus}} \\

\textbf{Randomized Variable}
& $C_r$ & $\Delta$C & $q$
& $C_r$ & $\Delta$C & $q$
& $C_r$ & $\Delta$C & $q$ \\
\hline

\textit{Reference}
& \multicolumn{3}{c|}{$C_0=0.7294$}
& \multicolumn{3}{c|}{$C_0=0.7280$}
& \multicolumn{3}{c}{$C_0=0.7402$} \\
\hline

Capsular Penetration
& 0.7493 & \makecell[tc]{$-0.0198$\\[-2pt]{\scriptsize[-0.0454,\ 0.0031]}} & 0.185
& 0.7424 & \makecell[tc]{$-0.0144$\\[-2pt]{\scriptsize[-0.0372,\ 0.0093]}} & 0.228
& 0.7527 & \makecell[tc]{$-0.0126$\\[-2pt]{\scriptsize[-0.0297,\ 0.0030]}} & 0.138 \\

Tertiary Gleason
& 0.7276 & \makecell[tc]{$0.0018$\\[-2pt]{\scriptsize[-0.0049,\ 0.0092]}} & 0.749
& 0.7161 & \makecell[tc]{$0.0119$\\[-2pt]{\scriptsize[-0.0016,\ 0.0294]}} & 0.137
& 0.7402 & \makecell[tc]{$0.0000$\\[-2pt]{\scriptsize[0,\ 0]}} & -- \\

Pre-operative PSA
& 0.7282 & \makecell[tc]{$0.0012$\\[-2pt]{\scriptsize[-0.0074,\ 0.0118]}} & 0.834
& 0.7163 & \makecell[tc]{$0.0117$\\[-2pt]{\scriptsize[0.0006,\ 0.0249]}} & 0.080
& 0.7274 & \makecell[tc]{$0.0128$\\[-2pt]{\scriptsize[-0.0004,\ 0.0299]}} & 0.116 \\

pT Stage
& 0.7337 & \makecell[tc]{$-0.0043$\\[-2pt]{\scriptsize[-0.0289,\ 0.0173]}} & 0.778
& 0.7078 & \makecell[tc]{$0.0203$\\[-2pt]{\scriptsize[-0.0097,\ 0.0501]}} & 0.201
& 0.7333 & \makecell[tc]{$0.0069$\\[-2pt]{\scriptsize[-0.0065,\ 0.0208]}} & 0.309 \\

Lymphovascular Invasion
& 0.7175 & \makecell[tc]{$0.0119$\\[-2pt]{\scriptsize[-0.0403,\ 0.0638]}} & 0.749
& 0.7045 & \makecell[tc]{$0.0235$\\[-2pt]{\scriptsize[-0.0393,\ 0.0865]}} & 0.453
& 0.6938 & \makecell[tc]{$0.0464$\\[-2pt]{\scriptsize[-0.0196,\ 0.1186]}} & 0.191 \\

Seminal Vesicle Invasion
& 0.7058 & \makecell[tc]{$0.0237$\\[-2pt]{\scriptsize[-0.0176,\ 0.0665]}} & 0.398
& 0.6853 & \makecell[tc]{$0.0427$\\[-2pt]{\scriptsize[0.0082,\ 0.0807]}} & \textbf{0.046}
& 0.6802 & \makecell[tc]{$0.0599$\\[-2pt]{\scriptsize[-0.0047,\ 0.1292]}} & 0.116 \\

Positive Surgical Margins
& 0.7049 & \makecell[tc]{$0.0245$\\[-2pt]{\scriptsize[-0.0037,\ 0.0570]}} & 0.185
& 0.6788 & \makecell[tc]{$0.0492$\\[-2pt]{\scriptsize[-0.0027,\ 0.1096]}} & 0.111
& 0.6885 & \makecell[tc]{$0.0516$\\[-2pt]{\scriptsize[-0.0087,\ 0.1132]}} & 0.116 \\

Secondary Gleason
& 0.6974 & \makecell[tc]{$0.0320$\\[-2pt]{\scriptsize[-0.0083,\ 0.0718]}} & 0.191
& 0.6883 & \makecell[tc]{$0.0397$\\[-2pt]{\scriptsize[0.0060,\ 0.0782]}} & \textbf{0.047}
& 0.6932 & \makecell[tc]{$0.0470$\\[-2pt]{\scriptsize[-0.0046,\ 0.1036]}} & 0.116 \\

All Gleason
& 0.6474 & \makecell[tc]{$0.0820$\\[-2pt]{\scriptsize[0.0055,\ 0.1615]}} & 0.110
& 0.6650 & \makecell[tc]{$0.0630$\\[-2pt]{\scriptsize[0.0125,\ 0.1167]}} & \textbf{0.046}
& 0.6790 & \makecell[tc]{$0.0612$\\[-2pt]{\scriptsize[-0.0082,\ 0.1318]}} & 0.116 \\

Primary Gleason
& 0.6405 & \makecell[tc]{$0.0889$\\[-2pt]{\scriptsize[0.0278,\ 0.1528]}} & \textbf{0.025}
& 0.6950 & \makecell[tc]{$0.0330$\\[-2pt]{\scriptsize[-0.0036,\ 0.0741]}} & 0.121
& 0.6760 & \makecell[tc]{$0.0642$\\[-2pt]{\scriptsize[0.0208,\ 0.1128]}} & \textbf{0.010} \\

Positive Lymph Nodes
& 0.6002 & \makecell[tc]{$0.1292$\\[-2pt]{\scriptsize[0.0552,\ 0.2057]}} & \textbf{0.012}
& 0.6264 & \makecell[tc]{$0.1017$\\[-2pt]{\scriptsize[0.0409,\ 0.1677]}} & \textbf{0.012}
& 0.7402 & \makecell[tc]{$0.0000$\\[-2pt]{\scriptsize[0,\ 0]}} & -- \\

\hline
All Pathology Parameters
& 0.5504 & \makecell[tc]{$0.1790$\\[-2pt]{\scriptsize[0.0501,\ 0.3066]}} & \textbf{0.033}
& 0.5168 & \makecell[tc]{$0.2112$\\[-2pt]{\scriptsize[0.0708,\ 0.3505]}} & \textbf{0.028}
& 0.4883 & \makecell[tc]{$0.2519$\\[-2pt]{\scriptsize[0.1239,\ 0.3733]}} & \textbf{0.001} \\
\hline

\end{tabular}
}

\vspace{0.2cm}
\footnotesize
\textit{Note:} $\Delta C > 0$ indicates reduced performance under the analysis condition. Features with $\Delta C = 0$ (predictions unaffected by randomization) are not testable and are marked --.
\end{table}

%% file: tables/feature_importance_multimodal_models.tex
\begin{table}[H]
\centering
\caption{Overview of performance per multimodal model when pathology-derived variables are removed. Models were bootstrapped (10,000 samples). Baseline performance is denoted by $C_0$ and post-variable removal performance by $C_r$. $\Delta$C-index ($C_0 - C_r$) is reported with 95\% bootstrap confidence intervals.}
\vspace{0.2cm}
\label{tab:feature_importance_MM}

\scriptsize
\renewcommand{\arraystretch}{1.15}

\begin{tabular}{l|c|c|c|c|c}
\hline
\textbf{Team}
& \textbf{Modalities}
& \textbf{$C_0$}
& \textbf{$C_r$}
& \textbf{$\Delta$C [95\% CI]}
& \textbf{$p$} \\
\hline

IU CompPath
& Clin + MRI + WSI
& 0.7197
& 0.6869
& \makecell[tc]{$0.0328$\\[-2pt]{\scriptsize[-0.0006,\ 0.0693]}}
& 0.056 \\

VCMI
& Clin + MRI + WSI
& 0.7153
& 0.6762
& \makecell[tc]{$0.0391$\\[-2pt]{\scriptsize[-0.0105,\ 0.0949]}}
& 0.124 \\

OHSU-Cedar
& Clin + WSI
& 0.6885
& 0.6881
& \makecell[tc]{$0.0004$\\[-2pt]{\scriptsize[-0.1027,\ 0.1053]}}
& 0.993 \\
\hline

\end{tabular}

\vspace{0.2cm}
\footnotesize
\textit{Note:} $\Delta C > 0$ indicates reduced performance under the analysis condition.
\end{table}

%% file: tables/modality_importance_VCMI.tex
\begin{table}[H]
\centering
\caption{Ablation analysis for VCMI under different modality and clinical feature configurations. Baseline (full model) performance is denoted by $C_0$ and the ablated performance by $C_r$. $\Delta$C-index ($C_0 - C_r$) is reported as the difference between baseline and ablation performance. Benjamini--Hochberg (BH) FDR-adjusted two-sided $p$-values are reported as $q$.}
\vspace{0.2cm}
\label{tab:ablation_vcmi}

\scriptsize
\renewcommand{\arraystretch}{1.15}

\begin{tabular}{l|c|c|c}
\hline
\textbf{Configuration}
& \textbf{$C_r$}
& \textbf{$\Delta$C-index}
& \textbf{$q$} \\
\hline

\textit{WSI + MRI + Clinical (full) - Reference}& $C_0 = 0.7153$
& -& -\\
\hline

WSI + Clinical (full)
& 0.7076
& 0.0077
& 0.5435 \\

WSI + Clinical (filtered)
& 0.6981
& 0.0172
& 0.5435 \\

WSI + MRI
& 0.6902
& 0.0251
& 0.4254 \\

Clinical only (full)
& 0.6831
& 0.0322
& 0.4402 \\

WSI + MRI + Clinical (filtered)
& 0.6762
& 0.0391
& 0.2488 \\

MRI + Clinical (full)
& 0.6687
& 0.0466
& 0.4254 \\

WSI only
& 0.6460
& 0.0693
& \textbf{0.0355} \\

Clinical only (filtered)
& 0.5062
& 0.2091
& \textbf{0.0073} \\

MRI + Clinical (filtered)
& 0.5010
& 0.2143
& \textbf{0.0073} \\

MRI only
& 0.4595
& 0.2558
& \textbf{0.0040} \\
\hline

\end{tabular}

\vspace{0.2cm}
\footnotesize
\textit{Note:} $\Delta C > 0$ indicates reduced performance relative to the baseline model.
$q$ denotes BH FDR-adjusted two-sided $p$-values; values in bold denote $q < 0.05$.
\end{table}

%% file: sections/discussion.tex
\section{Discussion}
\label{sec:discussion}

CHIMERA was developed to evaluate multimodal prognostic AI in a setting that reflects the integrated nature of clinical risk assessment after radical prostatectomy. Established clinical risk models already integrate information from multiple diagnostic processes implicitly, through the combined judgment of pathologists, radiologists, and treating clinicians. The next step is to evaluate whether AI models can perform this integration directly from raw multimodal data. By providing a fully curated and matched dataset integrating patient characteristics, clinician-derived variables, mpMRI, and WSIs from the same patients under a fixed evaluation protocol, with all algorithms assessed under the same per-case time limit and GPU memory budget, CHIMERA enables direct comparison of these data sources for prognostic modeling. The challenge results show that unimodal models built on clinician-derived variables outperformed multimodal models on the leaderboard, but post-hoc analyses indicate this asymmetry reflects the compressed expert signal embedded in those variables rather than a fundamental limit of multimodal learning.

The challenge results showed that unimodal submissions using patient characteristics and clinician-derived variables achieved higher C-indices than the multimodal submissions, which warrants further investigation. One explanation lies in the origin of the clinician-derived data. The non-imaging modality contained a curated dataset of manually extracted variables from the pathology report alongside patient characteristics such as age and PSA. Obtaining these variables requires an experienced pathologist to review and report on the resection specimen, followed by manual extraction from free-text pathology reports by a trained researcher. This is a labor-intensive process, and the resulting structured representation is a compressed summary of the information contained in the pathology imaging data. We hypothesized that this process yields a compact, high-quality feature set that, under the constraint of a small training cohort, encoded a stronger prognostic signal than raw imaging data alone. We therefore performed a post-hoc analysis with two aims: to quantify the dependence of unimodal models on clinician-derived variables, and to test whether multimodal models can recover equivalent signal from raw imaging when those variables are withheld.

Permuting pathology-derived variables across patients in the unimodal clinical models reduced performance to near-random levels ($C \approx$ 0.50, $q <$ 0.05), indicating that predictive capacity was driven primarily by these structured pathology-derived variables rather than by patient characteristics such as age and PSA. In contrast, multimodal models retrained without pathology-derived variables retained C-indices between 0.6762 and 0.6881, with no statistically significant decline relative to their full-modality counterparts. Although these values fall below the full unimodal leaderboard performance, they remain well above chance and suggest that multimodal models can extract prognostically relevant information directly from WSIs and MRI, partially compensating for the absence of manually curated pathological variables.

The higher performance of unimodal models on the leaderboard therefore partly reflects an asymmetry in the information available to each model type: unimodal models benefit from clinician-derived features that are themselves multimodal in origin, while multimodal models must learn equivalent representations from raw imaging data under the constraint of a small training cohort. 

This asymmetry was illustrated by the top-ranked team, whose model combining patient characteristics, clinician-derived variables, and WSIs achieved an internal validation C-index of 0.8770 but dropped to 0.6446 on the validation leaderboard, consistent with overfitting. The team's submitted model, using only patient characteristics and clinician-derived variables, generalized better. This has practical implications: in clinical settings where complete and structured pathology reports are available, such variables remain strong prognostic baselines. However, as cohort sizes grow into the thousands, manual extraction becomes infeasible; multimodal models offer a potential route toward reducing dependency on manual feature extraction and enabling more scalable prognostic modeling. This trajectory is supported by imaging-based benchmarks such as LEOPARD, where pathology-only AI models trained on larger datasets performed competitively with models requiring clinician-derived variables for recurrence prediction \cite{grisi2026leopard, faryna2026leopard}.

These findings are broadly consistent with the published landscape for BCR prediction after radical prostatectomy. The top test C-index of 0.7402 achieved in CHIMERA falls below the range of established post-operative clinical nomograms, including CAPRA-S (C-index = 0.77 in the original validation cohort) \cite{Cooperberg2011} and the Stephenson MSKCC nomogram (C-index range 0.79 to 0.81) \cite{Stephenson2005}. Direct numerical comparison should be interpreted cautiously, as reported performance is influenced by differences in cohort composition, sample size, follow-up duration, censoring rate, BCR definition, and validation setting. For example, multimodal fusion models have reported higher internal performance, such as the C-index of 0.860 reported by Hu et al. in a single-center cohort of 363 patients \cite{Hu2024}, but such single-center internal validation results are not directly comparable to performance on a held-out challenge test set.

Most established clinical nomograms rely on post-operative pathological variables such as Gleason grade, surgical margin status, capsular penetration, seminal vesicle invasion, and lymph node involvement: clinically meaningful, interpretable, strongly prognostic, and manually derived from the resection specimen. The CHIMERA post-hoc analyses showed that multimodal models retain prognostic performance when these variables are withheld, suggesting that raw WSI and MRI data contain recoverable signal overlapping with, and potentially complementing, established pathological summaries. CHIMERA does not replace clinical nomograms but provides a standardized setting to study how data-driven multimodal models relate to established risk models and to the expert-derived variables on which they depend.

To further characterize modality contributions within a single multimodal framework, an additional ablation analysis was performed using the VCMI model across all combinations of WSI, MRI, Clin\_full, and Clin\_filtered. The relationship between modalities was neither additive nor straightforward. Clin\_full alone achieved near-reference performance, consistent with the leaderboard results for the VCMI model. MRI in isolation performed close to chance, but combined with WSI it recovered substantial performance; adding MRI to Clin\_full, however, did not improve over Clin\_full alone. Clin\_filtered was similarly weak in isolation but additive when combined with WSI. These patterns suggest that the contribution of a modality depends on the information already available to the model rather than its isolated predictive capacity, and that MRI and patient characteristics contribute less to predictive performance when stronger clinician-derived pathological variables are already available, consistent with partial redundancy between raw imaging, patient characteristics, and clinician-derived pathology summaries. These observations should be interpreted cautiously. All ablations were conducted within a single multimodal architecture trained on 95 cases, meaning observed differences likely reflect both intrinsic modality information content and the sensitivity of the VCMI gated fusion mechanism to modality configuration and model complexity. Adding modalities in this regime increases capacity and overfitting risk without guaranteed gains in generalization. The finding that MRI improved WSI-based performance but not Clin\_full-based performance should therefore not be read as a general conclusion about MRI for BCR prediction, but as an illustration that modality interactions are context- and architecture-dependent, motivating standardized benchmarks such as CHIMERA where modality combinations can be compared under fixed data, preprocessing, and evaluation conditions.

The primary limitation of this challenge was the dataset size. Requiring complete mpMRI, pathology, and clinical data for all patients ensured comparability across submissions but imposed a strict inclusion criterion that limited the eligible patient pool. This trade-off is inherent to fully paired multimodal benchmarking: controlling data conditions enables standardized comparison across modalities and teams, but reduces cohort size and does not reflect the incomplete modality availability common in real-world clinical data. Consequently, the finding that multimodal submissions did not outperform models using structured clinical and pathology-derived variables should not be interpreted as evidence against multimodal prognostic modeling. Rather, it reflects the difficulty of training and evaluating high-dimensional multimodal models on fully paired cohorts under limited-data conditions.

This limitation is linked to origin of the current dataset. Although CHIMERA represents, to our knowledge, the largest fully paired multimodal challenge dataset for BCR prediction after radical prostatectomy, many initially considered patients had to be excluded to obtain a fully curated cohort with complete mpMRI, pathology, and clinical data. Expanding CHIMERA to additional institutions would increase sample size, introduce diversity in pathology and radiological scanners and in acquisition protocols, broaden the range of pathology processing and reporting practices, and improve generalizability. Larger datasets would also allow the ablation analyses performed here to yield more robust conclusions about modality contributions.

In a multicenter context, where modality availability inevitably varies across institutions, accommodating missing modalities becomes both a practical necessity and an opportunity to open the challenge to a broader range of modeling approaches. Future iterations could also leverage existing larger unimodal cohorts. Previous prostate histopathology challenges have demonstrated that AI models can learn prognostic information from WSIs beyond conventional Gleason-based assessment \cite{Bulten2022, grisi2026leopard, faryna2026leopard}. In CHIMERA, established clinicopathological predictors were included to provide strong reference variables and to enable participants to investigate whether additional prognostic signals could be learned from the pathology images. However, this was not clearly demonstrated in the current challenge, as multimodal models did not outperform models using clinician-derived pathology variables. This highlights the difficulty of learning both modality-specific representations and multimodal fusion strategies from limited paired data. A promising direction is therefore to decouple these two learning problems: large pathology and radiology datasets can be used to learn robust unimodal representations through supervised or self-supervised pretraining, after which the CHIMERA cohort can be used for multimodal fine-tuning and fusion. This strategy may be better suited to the limited availability of fully paired multimodal data than training all components directly on the challenge cohort.

A second limitation concerns the evaluation framework. The C-index was chosen as the primary metric because it accommodates right-censored observations, is robust to heterogeneous follow-up durations, and is model-agnostic, accepting predicted risk scores such as hazard scores, survival probabilities, or risk logits. These properties make it appropriate for prognostic modeling and consistent with established post-operative risk models such as CAPRA-S \cite{Cooperberg2011}. However, the C-index captures only the relative ordering of predicted risks and does not assess whether predicted risks are well calibrated or whether predicted event times are clinically meaningful. This is particularly relevant because many submitted models produced logits or relative risk scores, which support concordance-based ranking but cannot be directly interpreted as patient-specific time-to-event estimates without additional calibration. Due to the continuous nature of the reference and the pairwise evaluation it entails, a case-level analysis was not included. The high censoring rate in prostate cancer BCR cohorts ($\approx$ 70\%) further reduces the number of observed events and informative comparisons available for both training and evaluation. Future CHIMERA editions could complement the C-index with calibration assessment, time-dependent AUC, and Brier score. These additions would allow models to be evaluated beyond their ability to rank patients by recurrence risk, better aligning model evaluation with clinical decision-making and enabling more direct comparison with established clinical risk scores.

Several of these limitations here are directly addressed in a follow up challenge, CHIMERA-Agent (\url{https://chimera-agent.grand-challenge.org/}), which extends this benchmark towards agentic multimodal reasoning across the prostate cancer patient journey under realistic clinical conditions, including missing modalities. The CHIMERA training data will remain publicly available for the foreseeable future. 

%% file: sections/conclusion.tex
\section{Conclusion}
\label{sec:conclusion}

The CHIMERA challenge introduces the first urological multimodal benchmark for prostate cancer prognostic modeling, providing a fully curated dataset from experienced urological centers in the Netherlands comprising clinical data in the form of patient characteristics and clinician-derived variables, multiparametric MRI, and digitized H\&E whole slide prostatectomy images. Six teams submitted methods covering a wide range of modality combinations and fusion strategies, enabling a structured comparison of multimodal deep learning approaches under controlled and reproducible conditions.

Challenge results showed that pathological variables currently drive predictive performance most strongly, and that multimodal models did not consistently outperform unimodal clinical models on the held-out test set. Post-hoc analyses showed that when clinician-derived variables were withheld, unimodal models lost nearly all predictive performance, whereas multimodal models retained substantially more. This suggests that multimodal approaches are less dependent on expert feature engineering. The implication for the future is direct: as datasets scale and complete pathological annotation is not always available, multimodal models may be better positioned to generalize and improve without requiring extensive preprocessing pipelines.

Multimodal prognostic modeling in prostate cancer is a rapidly developing field, and as increasingly sophisticated deep learning approaches are proposed, the need for fair and structured evaluation grows accordingly. CHIMERA provides the first benchmark infrastructure to support this. Future iterations should expand sample size, incorporate additional centers to better capture population heterogeneity, introduce evaluation metrics beyond the concordance index, and handle missing data in ways that more faithfully reflect clinical reality. CHIMERA provides the foundation on which this ongoing effort can build.

%% file: sections/credit_section.tex
\section*{CRediT authorship contribution statement}

\textbf{Robert N. Spaans:} Conceptualization, Data curation, Formal analysis, Investigation, Methodology, Project administration, Software, Validation, Visualization, Writing -- original draft, Writing -- review and editing.
\textbf{Catherine Chia:} Conceptualization, Project administration, Writing -- review and editing.
\textbf{Tongjie Wang:} Writing -- review and editing.
\textbf{Adam Kowalewski:} Data curation, Writing -- review and editing.
\textbf{Parandzem Khachatryan:} Data curation, Writing -- review and editing.
\textbf{Domingos Oliveira:} Data curation, Writing -- review and editing.
\textbf{Khrystyna Faryna:} Conceptualization, Software, Writing -- review and editing.
\textbf{Jean-Paul A. van Basten:} Resources, Supervision, Writing -- review and editing.
\textbf{Geert Litjens:} Conceptualization, Data curation, Resources, Software, Supervision, Writing -- review and editing.
\textbf{Nadieh Khalili:} Conceptualization, Funding acquisition, Methodology, Project administration, Resources, Software, Supervision, Writing -- review and editing.
\textbf{Noorul Wahab, Ethar Alzaid, Jiaqi Lv, Adam Shephard, Shan E Ahmed Raza, Yeonwoo Seo, Emily Xie, Jun Ma, Bo Wang:} Methodology, Writing -- review and editing.
\textbf{Felipe C. M. R. Coutinho, Pedro Vitor Lima, Leonardo M. Ferreira, In\^{e}s P. Machado, Jaime S. Cardoso, Mustafa Akur, Selim Sevim, Suhang You, Carla Pitarch-Abaigar, Sanket Kachole, Sumedh Sonawane, Spyridon Bakas:} Methodology, Software, Writing -- review and editing.
\textbf{Xulin Chen, Junzhou Huang:} Methodology, Writing -- review and editing.

%% file: sections/consortium.tex
\section*{CHIMERA Consortium}
\noindent
\textbf{Noorul Wahab}$^{k}$,
\textbf{Ethar Alzaid}$^{k}$,
\textbf{Jiaqi Lv}$^{k}$,
\textbf{Adam Shephard}$^{k}$,
\textbf{Shan E Ahmed Raza}$^{k}$,
\textbf{Yeonwoo Seo}$^{l}$,
\textbf{Emily Xie}$^{m,n}$,
\textbf{Jun Ma}$^{o}$,
\textbf{Bo Wang}$^{n,o,p,q,r}$,
\textbf{Felipe C. M. R. Coutinho}$^{s}$,
\textbf{Pedro Vitor Lima}$^{s}$,
\textbf{Leonardo M. Ferreira}$^{s}$,
\textbf{In\^{e}s P. Machado}$^{t}$,
\textbf{Jaime S. Cardoso}$^{s}$,
\textbf{Mustafa Akur}$^{u}$,
\textbf{Selim Sevim}$^{u}$,
\textbf{Suhang You}$^{v,w}$,
\textbf{Carla Pitarch-Abaigar}$^{v,w}$,
\textbf{Sanket Kachole}$^{v,w}$,
\textbf{Sumedh Sonawane}$^{x}$,
\textbf{Spyridon Bakas}$^{v,w,x,y,z}$,
\textbf{Xulin Chen}$^{aa}$,
\textbf{Junzhou Huang}$^{aa}$
\bigskip

\noindent $^{k}$Tissue Image Analytics Centre, Department of Computer Science, University of Warwick, Coventry, United Kingdom.
$^{l}$Department of Computer Science and Engineering, Korea University, Seoul, Republic of Korea.
$^{m}$Department of Mathematics, University of California San Diego, La Jolla, CA, USA.
$^{n}$Vector Institute for Artificial Intelligence, Toronto, Canada.
$^{o}$AI Hub, University Health Network, Toronto, Canada.
$^{p}$Peter Munk Cardiac Centre, University Health Network, Toronto, Canada.
$^{q}$Department of Computer Science, University of Toronto, Toronto, Canada.
$^{r}$Department of Laboratory Medicine and Pathobiology, University of Toronto, Toronto, Canada.
$^{s}$INESC TEC, Faculdade de Engenharia, Universidade do Porto, Porto, Portugal.
$^{t}$Department of Oncology, University of Cambridge, Cambridge, United Kingdom.
$^{u}$Cancer Early Detection Advanced Research Center (CEDAR), Knight Cancer Institute, Oregon Health and Science University, Portland, OR, USA.
$^{v}$Division of Computational Pathology, Department of Pathology and Laboratory Medicine, Indiana University School of Medicine, Indianapolis, IN, USA.
$^{w}$Indiana University Melvin and Bren Simon Comprehensive Cancer Center, Indianapolis, IN, USA.
$^{x}$Luddy School of Informatics, Computing, and Engineering, Indiana University, Bloomington, IN, USA.
$^{y}$Department of Radiology and Imaging Sciences, Indiana University School of Medicine, Indianapolis, IN, USA.
$^{z}$Department of Biostatistics and Health Data Science, Indiana University School of Medicine, Indianapolis, IN, USA.
$^{aa}$Department of Computer Science and Engineering, University of Texas at Arlington, Arlington, TX, USA.

%% file: sections/supplementary.tex
\section{Supplementary Material}
\label{sec:supplementary}

\renewcommand{\thesubsection}{\Alph{subsection}}
\setcounter{subsection}{0}

\input{supplementary_data/challenge_organizers.tex}

% --- Supplementary figures ---
\newpage
\subsection{Supplementary Figures}
\renewcommand{\figurename}{Supplementary Figure}
\renewcommand{\thefigure}{S\arabic{figure}}
\renewcommand{\theHfigure}{supp.\arabic{figure}}
\setcounter{figure}{0}

\begin{figure}[!ht]
    \centering
    \includegraphics[width=\linewidth]{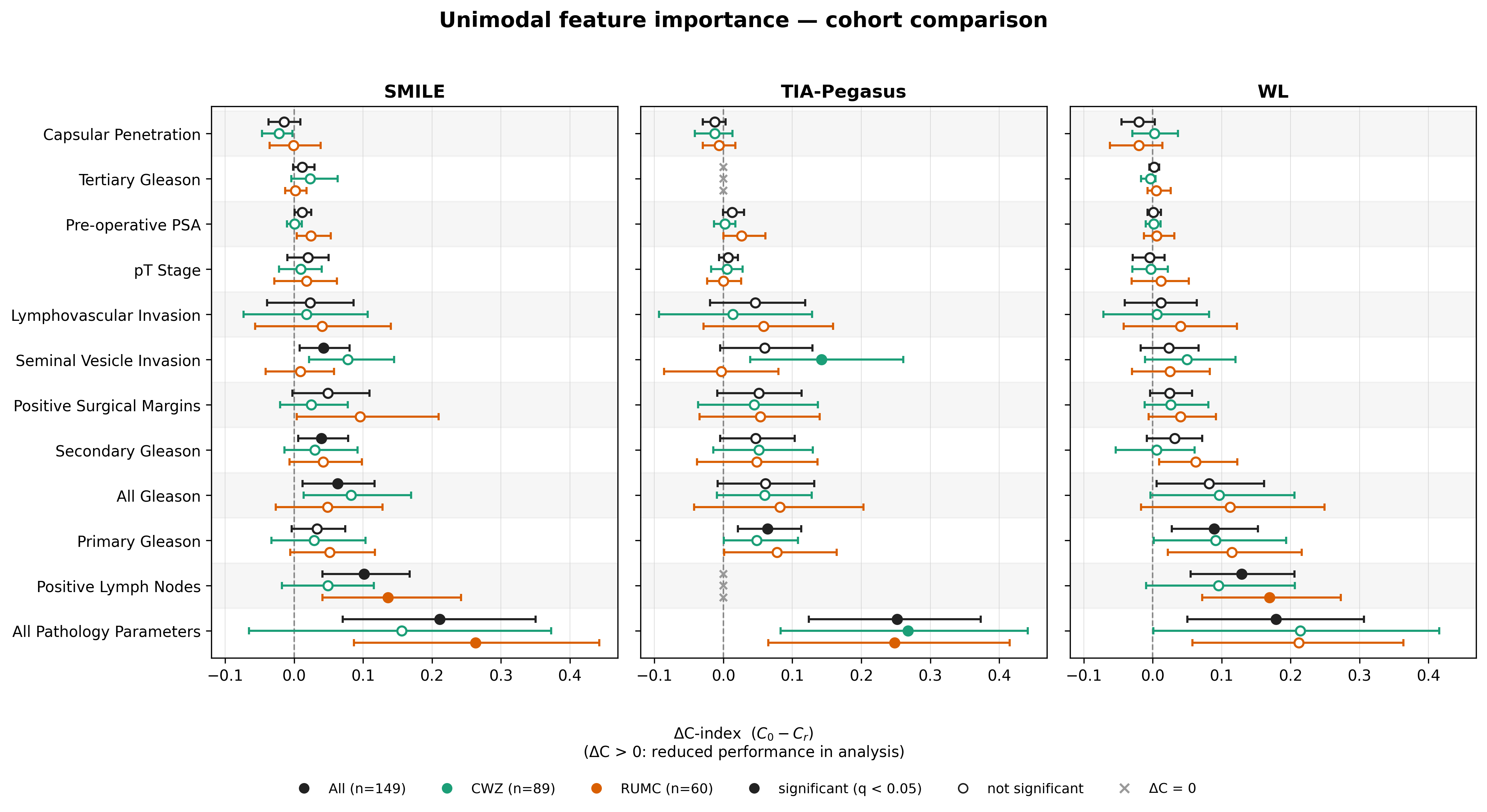}
    \caption{Forest plot of the performance per unimodal team for randomized variables per cohort.}
    \label{sup_fig:forest_unimodal_feature_importance}
\end{figure}

\begin{figure}[!ht]
    \centering
    \includegraphics[width=1\linewidth]{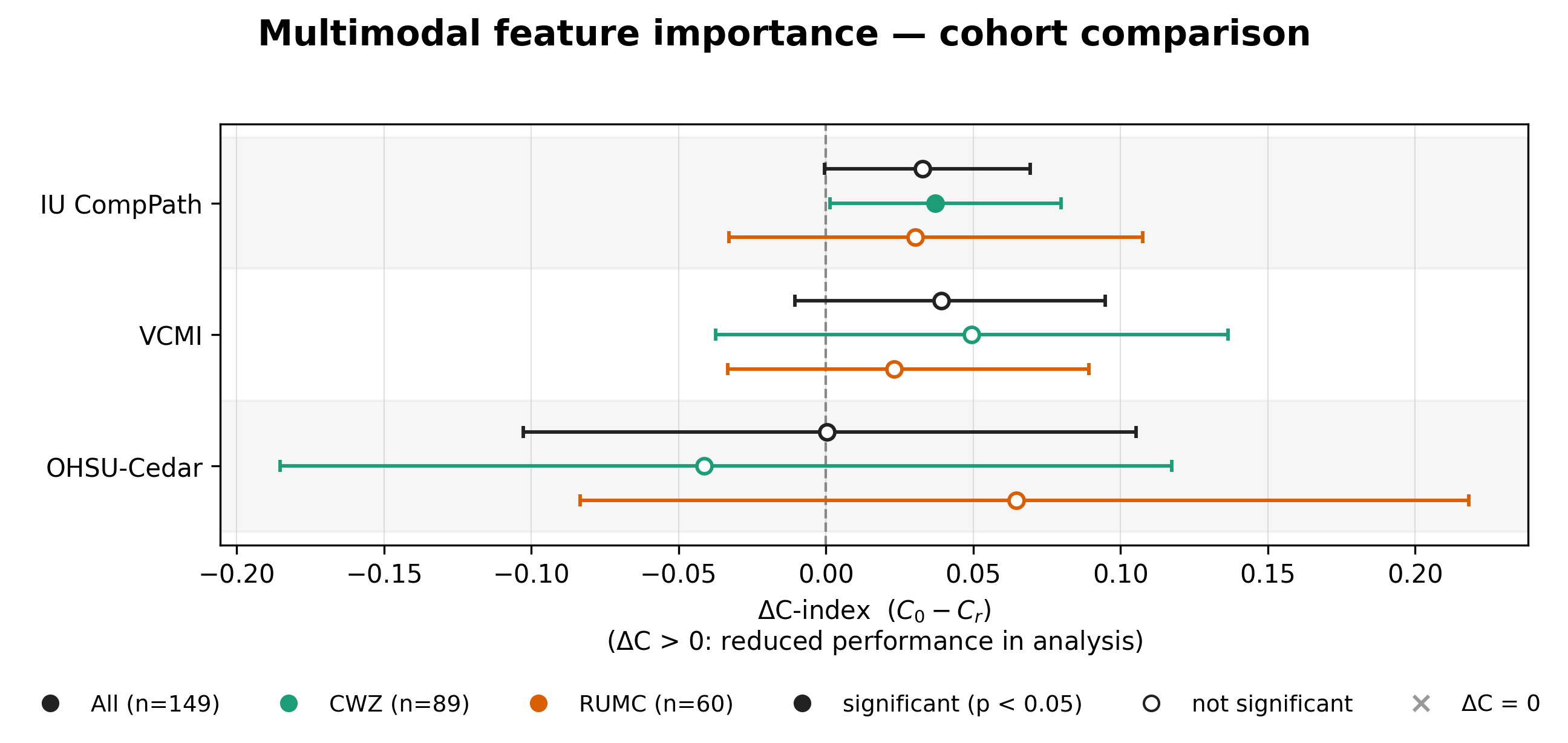}
    \caption{Forest plot of the performance per multimodal team when pathology-derived variables were removed, split per origin center of the data.}
    \label{sup_fig:forest_multimodal_feature_importance}
\end{figure}

\begin{figure}[!ht]
    \centering
    \includegraphics[width=1\linewidth]{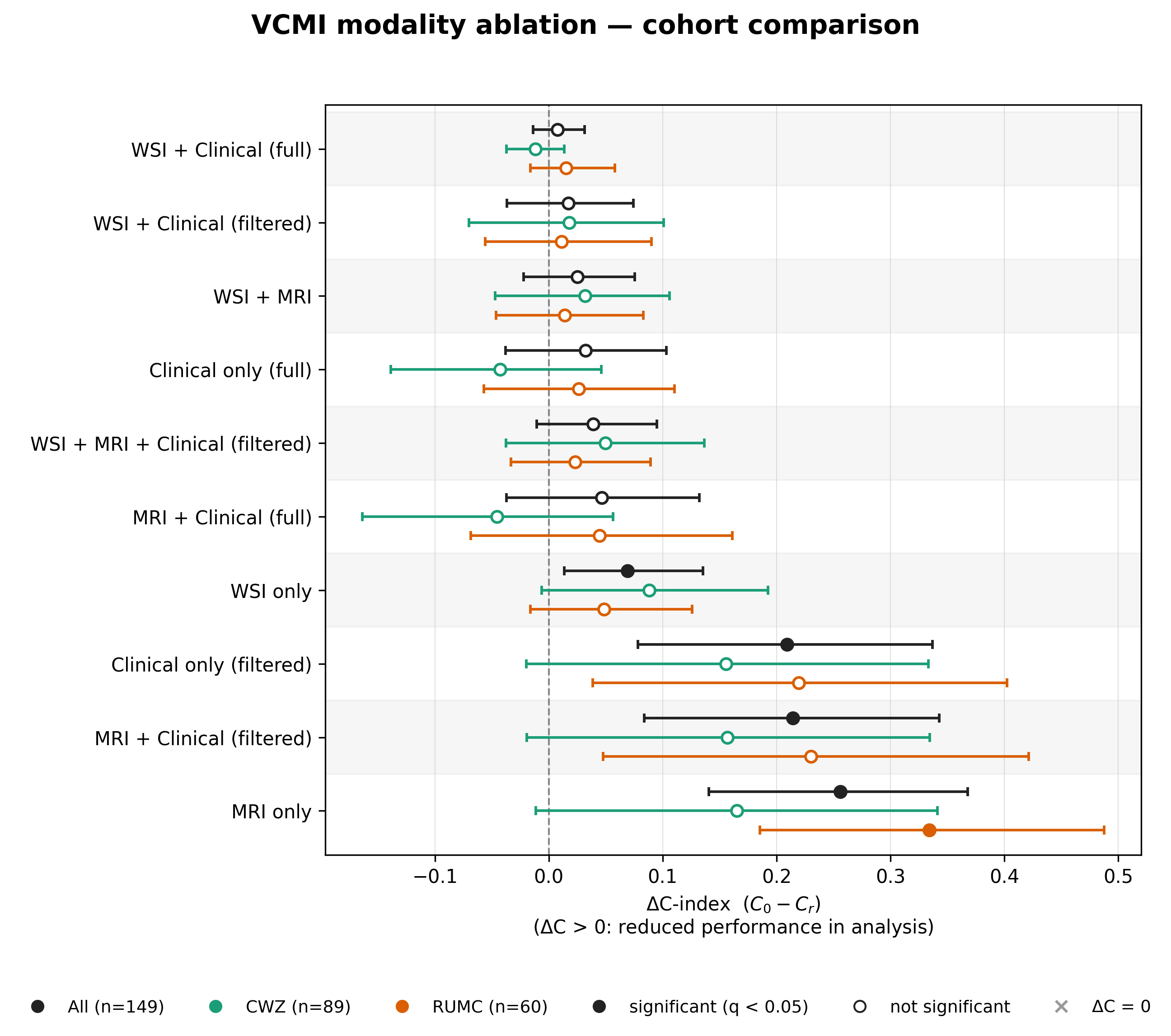}
    \caption{Forest plot of the ablation analysis for the VCMI model under different modality and clinical feature configurations per cohort}
    \label{sup_fig:forest_modality_ablation}
\end{figure}

%% file: supplementary_data/challenge_organizers.tex
\subsection{CHIMERA Challenge Organizers}
\label{sec:supp-organizers}

\vspace{0.5em}
\begin{flushleft}
Robert N. Spaans\textsuperscript{a,b,c},
Catherine Chia\textsuperscript{a,c,d,f},
Tongjie Wang\textsuperscript{a,e,f},
Farbod Khoraminia\textsuperscript{e},
Khrystyna Faryna\textsuperscript{a},
Maryam Mohammadlou\textsuperscript{a,g},
Tahlita Zuiverloon\textsuperscript{d},
Jean-Paul A. van Basten\textsuperscript{b},
Sita Vermeulen\textsuperscript{h},
Geert Litjens\textsuperscript{a,c},
Nadieh Khalili\textsuperscript{a}
\end{flushleft}

\vspace{0.5em}
\begin{flushleft}
\textsuperscript{a}Department of Pathology, Research Institute for Medical Innovation, Radboud University Medical Center, Nijmegen, The Netherlands\\
\textsuperscript{b}Department of Urology, Canisius Wilhelmina Hospital, Nijmegen, The Netherlands\\
\textsuperscript{c}Oncode Institute, Utrecht, The Netherlands\\
\textsuperscript{d}Department of Pathology, Erasmus University Medical Center, Rotterdam, The Netherlands\\
\textsuperscript{e}Department of Urology, Erasmus University Medical Center, Rotterdam, The Netherlands\\
\textsuperscript{f}Department of Dermatology, Erasmus University Medical Center, Rotterdam, The Netherlands\\
\textsuperscript{g}Prostate Cancer Research Center, Faculty of Medicine and Health Technology, Tampere University and Tays Cancer Centre, Tampere University Hospital, Tampere, Finland\\
\textsuperscript{h}Genetic Epidemiology, Radboud University Medical Center, Nijmegen, The Netherlands
\end{flushleft}